\documentstyle[prd,aps,floats]{revtex}

\begin{document}
\draft

\input epsf
\renewcommand{\topfraction}{0.8}
\twocolumn[\hsize\textwidth\columnwidth\hsize\csname
@twocolumnfalse\endcsname

\title{Protecting the Holographic Principle: Inflation}
\author{V\'{\i}ctor H. C\'{a}rdenas}
\address{Instituto de F\'{\i}sica, Universidad Cat\'{o}lica de Valpara\'{\i}so, Casilla 4059, Valpara\'{\i}so, Chile}

\maketitle

\begin{abstract}
A scenario where inflation emerges as a response to protect the
holographic principle is described. A two fluid model in a closed
universe inflation picture is assumed, and a possible explanation
for secondary exponential expansion phases as those currently
observed is given.
\end{abstract}

\pacs{PACS numbers: 98.80.Cq  \hfill UCV-IF 02/03 , gr-qc/0205070}

\vskip2pc]

\section{Introduction}

Although inflationary models have proven to be a very successful
tool for cosmology \cite{inf}, nobody knows why this period
started at early times. We only know how this finished, evolving
in a strong non-adiabatic and out-of-equilibrium phase, called
{\it reheating}. Because inflation started at times as early as
$10^{-35}\sec $, it is not difficult to believe that inflation is
a natural output coming from a Grand Unified Theory (GUT) holding
at Planck scale. So, we have no any idea of how inflation started,
because we cannot say anything about such a theory, .

However, 't Hooft \cite{thooft} proposed a crucial feature for
that theory: it has to be holographic. The idea of holography is
the following: if we want to reconcile quantum mechanics with
gravity, we have to assume that the observable degrees of freedom
of the universe are projections coming from a two-dimensional
surface, where the information is stored. Recently, there has been
a lot of interest in the relation between holography and string
theory \cite{string}.

The seminal work of Bekenstein \cite{bek} on the universal upper
bound on the entropy-energy ratio for bounded systems suggested
the idea of a holographic universal (HP) principle that can be
applied to cosmology. In this context, Fischler and Susskind
\cite{FS} proposed a holographic principle and studied its
consequences for the standard model of cosmology.

During the last years, there has been a lot of work that attempts
to refine the original proposal of Fischler and Susskind (FS).
This search has tried to come up with a HP which does not conflict
with inflation and cosmology in general. In \cite{holo} works on
this line can be found. Because the FS proposal was based on
adiabatic evolution and failed for closed
Friedman-Robertson-Walker `cosmologies', the interest was both
extend the study to non-adiabatic evolution (as reheating after
inflation) and try to include the closed case. The current
understanding based on these studies implies that this universal
principle does not constraint inflation.

In Ref.\cite{EL} the authors propose to replace the HP by the
generalized second law of thermodynamics. The same idea was
developed in \cite{KL}, where the authors explicitly discussed the
problems of the FS proposal applied to inflationary models.

The general objective of all these works is formulate the HP in
cosmology. Because the HP must be considered a universal principle
of a higher status than inflation, in this work I look for a
solution to the problem considering a different point of view: an
attempt to derive inflation from the HP. This is so because we
expect the HP, to come from a (so far unknown) GUT theory, and
then I will analyze if this new point of view may constraint any
inflationary model. From this declaration of principles, it is
clear that we do not find the current beliefs very informative
because inflation and the HP are considered (almost) independent
processes.

In an interesting work, Rama \cite{rama} proposed a scenario where
the holographic principle {\it ala}\ FS could be applied to a
closed universe. He showed that this is possible considering at
least one exotic field matter component with density pressure
ratio $w<-1/3$, and if the present value of density parameter
$\Omega$ is close to one. He also suggested that this component
could be realized by some form of `quintessence'\cite{QTS}.
However, as is stressed in \cite{KL}, this model is not correct
because it works under adiabatic evolution with a matter content
of a typical `inflationary' universe ($ -1<w<-1/3$), which is a
strong non-adiabatic process.

In this Letter, I look at the consequences of the FS proposal in
cosmology by considering the non-adiabatic process of reheating
after inflation. I discuss the case of a closed universe, which is
apparently weakly favored by observations \cite{closed}, with both
radiation and a real scalar field $\varphi $. At early times I
found that an imminent violation of the HP forces the system to
saturate it through a period of exponential growth of the scale
factor. This effect can be interpreted as a response to protect
the HP inside the causal volume associated. Also, I discuss the
possibility to explain additional exponential expansions like the
one observed now. Although it is possible to obtain an accelerated
phase that resembles the quintessential inflation picture
\cite{PV}, it is not due to a mechanism of saturation of the HP
{\it a la} Fishler-Susskind. The results of a numerical
integration of the evolution from inflation to the present are
shown in Figure 1. The key ingredient to have an additional phase
of accelerated expansion is reheating. The extremely efficient
transfer of energy from the scalar field to radiation enables us
to obtain a radiation-dominated phase with a vanishing small relic
of $\varphi $ fields. This relic does not confront observations
because the universe remains almost flat (until very small
redshift) and during matter domination, it is subdominant;
however, it can dominate the matter contributions for $z<0.5$. The
holographic principle of Fishler and Susskind is reviewed in
section II. I present the model in section III and develop the
formulae to treat reheating in section IV. The effect of
protection of the HP is discussed in section V, and the
possibilities to have another phase of accelerated expansion is
discussed in section VI. The paper ends with a discussion.

\section{The Fischler-Susskind Holographic Principle}

Let us assume a closed homogeneous isotropic universe with metric
\begin{equation}
ds^{2}=dt^{2}-a^{2}(t)(d\chi ^{2}+\sin ^{2}\chi d\Omega ^{2}),  \label{line}
\end{equation}
where $\chi $ is the azimuthal angle of $S_{3}$ and $\Omega $ is
the solid angle parametrizing the two-sphere at fixed $\chi $. The
particle horizon is
\begin{equation}
\chi _{H}=\int_{t_{i}}^{t}\frac{dt^{\prime }}{a(t^{\prime })},  \label{horiz}
\end{equation}
where $t_{i}$ is a reference initial time. Because integral
(\ref{horiz}) may diverge at small $t$, a natural choice is to
take $t_{i}=t_{pl}=1$ \cite{KL}. The angle $\chi _{H}$ determines
the coordinate size of the horizon, which defines a bounded area
and volume. Because the entropy density $\sigma \equiv (\rho
+p)/T$ is constant, the entropy area ratio gives
\begin{equation}
\frac{S}{A}=\sigma \frac{2\chi _{H}-\sin 2\chi _{H}}{4a^{2}(\chi _{H})\sin
^{2}(\chi _{H})}.  \label{s/a}
\end{equation}
As the universe evolves, ratio (\ref{s/a}) increases and the
system reaches a stage of saturation and later, a violation of
holographic principle \cite{FS}. For example, for a universe
filled with non-relativistic matter, $a=a_{\max }\sin ^{2}(\chi
_{H}/2)$ so for its maximal expansion $\chi _{H}=\pi $ ratio
(\ref{s/a}) becomes violated.

\section{The model}

I model the universe as filled with both, a single scalar field -
inflaton $\varphi $ - and a fluid of relativistic particles (or
radiation) with energy density $\rho _{m}$. Assuming a homogeneous
field with a slowly-time-dependent equation of state $p_{\varphi
}=w(t)\rho _{\varphi }$, where
\begin{equation}
w(t)=\frac{\dot{\varphi}^{2}-2V(\varphi )}{\dot{\varphi}^{2}+2V(\varphi )},
\label{ratio}
\end{equation}
we can write the Friedman equation as
\begin{equation}
H^{2}=H_{0}^{2}\left[ \Omega _{\varphi }\left( \frac{a_{0}}{a}\right)
^{3(1+w)}+\Omega _{m}\left( \frac{a_{0}}{a}\right) ^{4}-\Omega _{k}\left(
\frac{a_{0}}{a}\right) ^{2}\right] ,  \label{hubble}
\end{equation}
where $a_{0}$ and $H_{0}$ are the scale factor and the Hubble parameter in
an arbitrary reference time $t_{0}$ (often taken to be the present time).
Also, we have defined the following parameters: the density parameter $%
\Omega _{i}=\rho _{0i}/\rho _{c}$ for the $i$ component with
initial energy density $\rho _{0i}$; the critical density $\rho
_{c}=3H_{0}^{2}M_{p}^{2}/8\pi $, the equivalent energy density due
to
curvature $\rho _{k}=-3M_{p}^{2}/8\pi a^{2}$; and the Planck mass $%
M_{p}=1.2\times 10^{19}$GeV.

Because I am considering a pressure density ratio in the range $-1<w<-1/3$%
, this implies that at some time $t^{\ast }$ the first term in the
square brackets of (\ref{hubble}) will dominate over the other
contributions,
starting a period of `inflationary expansion' (if we also have $\ddot{a}>0$%
). This exponential expansion leads to a process of `flattening'
of the universe due to the screening of the curvature term. The
authors of \cite{KT} used the latter effect to obtain an apparent
spatially flat FRW universe by using a closed one.

If we extend the evolution to the present (i.e., at $t_{0}$), the
evolution of the system leads to a total density parameter
\[
\Omega =\sum_{i}\Omega _{i}=\Omega _{\varphi }+\Omega _{m}=1+\Omega
_{k}\simeq 1,
\]
because $\Omega _{k}=\rho _{0k}/\rho _{c}=3M_{p}^{2}/8\pi
a_{0}^{2}H_{0}^{2}\ll 1$ when we evaluate it today. Globally, the
effect agrees with the results of \cite{rama}, which to solve the
violation problem of the HP bound for a closed universe,
introduced an `exotic fluid' - as the inflaton here - as an extra
matter component. However, the presence of this component leads to
a period of non-adiabatic evolution \cite{KL}, so the equations of
motion must be corrected to consider entropy production during the
reheating process.

The simplest way to do that is to consider - as a first
approximation - the perturbative regime of reheating which was
described by adding a friction term to the inflaton equation of
motion \cite{REHEAT}. In this paper, I have taken into account the
most efficient energy transfer possible, based on the main results
of the modern reheating theory \cite{MthR}. A more realistic and
detailed approach based on this theory will be addressed elsewhere
\cite{next}.

\section{Particle and/or entropy production}

In Ref.\cite{KL} Kaloper and Linde concluded that the proposal of
Fischler and Susskind does not confront inflation, because it
eliminates the entropy produced inside the light cone during
reheating. This means that their HP is valid during adiabatic
expansion. To extend the analysis through the particle creation
process, we have to learn how to compute ratio (\ref {s/a}) in a
non-adiabatic stage, i.e., during reheating.

I consider the simplest model of reheating, which is based on the
Born approximation \cite{REHEAT}. Here, the inflaton evolves
according to
\begin{equation}
\ddot{\varphi}+3H\dot{\varphi}++V^{\prime }(\varphi )=-\Gamma \dot{\varphi},
\label{ieom}
\end{equation}
where $\Gamma $ is the rate of particle production, and the
evolution of the relativistic particles created is described by
\begin{equation}
\dot{\rho}_{m}+4H\rho _{m}=\Gamma \rho _{\varphi }.  \label{reom}
\end{equation}
Equations (\ref{ieom}) and (\ref{reom}) explicitly show the
non-adiabatic nature of the process. If, as usual, we define the
energy density and pressure by
\begin{equation}
\rho =\frac{1}{2}\dot{\varphi}^{2}+V(\varphi ),\ p=\frac{1}{2}\dot{\varphi}%
^{2}-V(\varphi ),  \label{eq2}
\end{equation}
we can write the equation of state during the rapid oscillations
phase of $\varphi $, i.e., $V^{\prime \prime }\gg H^{2}$ as a
temporal average during an oscillation by $\langle (\rho + p)/
\rho \rangle = w=(n-2)/(n+2)$ where $V(\varphi )=\varphi ^{n}$, so
\begin{equation}
\dot{\rho}_{\varphi }+3H\rho _{\varphi }(1+w)=-\Gamma \rho _{\varphi }.
\label{eq3}
\end{equation}
In the special case $w=0$ (a quadratic potential), this equation
describes the decay of a massive particle. A solution of this
equation is
\begin{equation}
\rho _{\varphi }=M^{4}\left( \frac{a_{i}}{a}\right) ^{3}\exp \left[ -\Gamma
\left( t-t_{i}\right) \right] ,  \label{eq4}
\end{equation}
where subscript $i$ indicates the epoch when the coherent
oscillations around the minimum of the potential $V(\varphi )$
begins, and $M^{4}$ is the vacuum energy at that time.

If the produced particles are thermalized, we can use the
expression for the entropy of radiation
\begin{equation}
S=\frac{2\pi ^{2}}{45}gT^{3}a^{3},  \label{eq6}
\end{equation}
which combined with the equality $\rho _{m}=\pi ^{2}gT^{4}/30$,
valid for relativistic particles and Eq.(\ref{eq6}), enables us to
write a relation between the energy density and the entropy
\begin{equation}
\rho _{m}=\frac{3}{4}\left( \frac{45}{2\pi ^{2}g}\right) ^{1/3}S^{4/3}a^{-4}.
\label{eq9}
\end{equation}
For $t<\Gamma ^{-1}$, the universe is dominated by $\varphi $
particles and, according to Eq.(\ref{hubble}) and (\ref{eq4}), it
evolves as a matter-dominated universe $a(t)\sim t^{2/3}$. An
approximated solution of (\ref {reom}) is
\[
\rho _{m}\simeq \frac{1}{10}M_{p}\Gamma M^{2}\left( \frac{a}{a_{i}}\right)
^{-4}\left[ \left( \frac{a}{a_{i}}\right) ^{5/2}-1\right] ,
\]
which implies that, initially $\rho _{m}$ increases from \ $0$ to $%
M_{p}^{2}\Gamma M^{2}$, and after that it decreases as $a^{-3/2}$
(see Figure 1). From (\ref{eq9}) we find that the entropy grows as
$S\propto a^{15/8}$. This means that $\sigma $ (which appears in
front of (\ref{s/a})) is no longer constant, and has to be
replaced by a function varying as $\sim $ $a^{15/8}$\cite
{turner85}.

\begin{figure}[t]
\centering \leavevmode\epsfysize=9cm \epsfbox{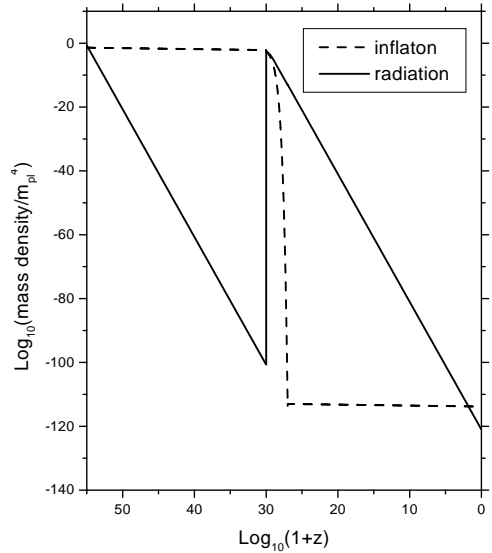}\\
\caption[fig1]{\label{fig1} Evolution in the model. Inflation
starts at $z=55$ and finishes at $z=30$. Note that at $z\sim 1$,
the inflaton field starts to dominate the matter content leading
to a new exponential expansion phase.}
\end{figure}

\section{Protecting the HP with Inflation}

Let us assume that the universe starts in a radiation dominated
era $\rho _{m}\sim \rho _{k}\gg \rho _{\varphi }$ at a time near
the beginning of inflation, say $10^{-35}\sec $. As was discussed
in Section II, such a universe will evolve towards an imminent
violation of the HP. In fact, during this phase Eq.(\ref{hubble})
can be written as
\begin{equation}
H^{2}=\frac{8\pi }{3M_{p}^{2}}\rho _{0m}\left( \frac{a_{0}}{a}\right) ^{4}-%
\frac{1}{a^{2}},  \label{initH}
\end{equation}
from which we obtain the solution
\begin{equation}
a(\chi _{H})=A\sin (\chi _{H}),  \label{sol1}
\end{equation}
where $A\equiv \sqrt{8\pi \rho _{0m}a_{0}^{4}/3M_{p}^{2}\text{ }}$. By
inserting (\ref{sol1}) in (\ref{s/a}), we find that the HP is violated as $%
\chi _{H}\rightarrow \pi $. Note that in this form, the bound
(\ref{s/a}) is violated also as $\chi _{H}\rightarrow 0$. This
particular problem is solved by the arguments displayed under
Eq.(\ref{horiz}). Because $a(\chi _{H}=0)=a_{p}\neq 0$, then it is
possible to write the solution (\ref{sol1}) as $a(\chi
_{H})=a_{p}+A\sin (\chi _{H})$, solving the HP bound (\ref{s/a})
as $\chi _{H}\rightarrow 0$. However, the violation in the future
($\chi _{H}\rightarrow \pi $) remains.

Now, let us follow the evolution of both (\ref{s/a}) and
(\ref{hubble}) during the transit from radiation domination to the
inflationary phase. If we assume at Planck time $t_{pl}$ that
$\rho _{m}\sim \rho _{k}$, then we can expect after certain time
say $t\simeq 10^{-35}\sec $, to enter an inflationary phase. In
fact, because $\rho _{\varphi }\sim a^{-3(1+w)}$ with $-1<w<-1/3$
decays more slowly than $\rho _{k}\sim a^{-2}$and $\rho _{m}\sim
a^{-4}$, the former dominates. If we consider the beginning of
inflation when $\rho _{\varphi }$ becomes comparable to $\rho
_{k}$, this implies that at the time $t\sim 10^{-37}\sec $ we have
\[
\rho _{\varphi }\simeq 10^{-2}(10^{-1})\rho _{k},
\]
for $w=-1$ ($-2/3$), respectively. After $\rho _{\varphi }$
becomes greater than $\rho _{k}$, this component dominates the
matter content in the universe and makes it inflationary in the
sense that $\rho _{\varphi }\gg \rho _{k},\rho _{m}$. Here the
Hubble parameter $H$ (\ref{hubble}) becomes nearly constant, and
the coordinate size of the horizon (\ref{horiz}) behaves like
\begin{equation}
\chi _{H}=\int_{t_{p}}^{t}\frac{dt^{\prime }}{a(t^{\prime })}\simeq
H^{-1}(e^{-Ht_{p}}-e^{-Ht}),  \label{hosat}
\end{equation}
reaching an asymptotic constant value proportional to $H^{-1}$.
Because the scale factor grows nearly exponentially, the HP bound
(\ref{s/a}) starts to decrease, avoiding the violation in the
future. A numerical integration of the system is shown in Fig. 2.
In this way, the appearance of inflation saves the violation of
the HP in the future. For example, in the $w=-1$
case, the scale factor can be written as $a(\chi _{H})\simeq \sqrt{3/8\pi }%
\left[ \sin (\pi /2-\chi _{H})\right] ^{-2}$, which behaves much
like an exponential growth. The early radiation dominated phase is
not necessary at all to demonstrate the role of the HP in
inflating the universe. In fact, we can start the universe with a
bound $S/A$ saturated at the Planck era; this time with $\rho
_{\varphi }\geq \rho _{k},\rho _{m}$.

\begin{figure}[t]
\centering \leavevmode\epsfysize=9cm \epsfbox{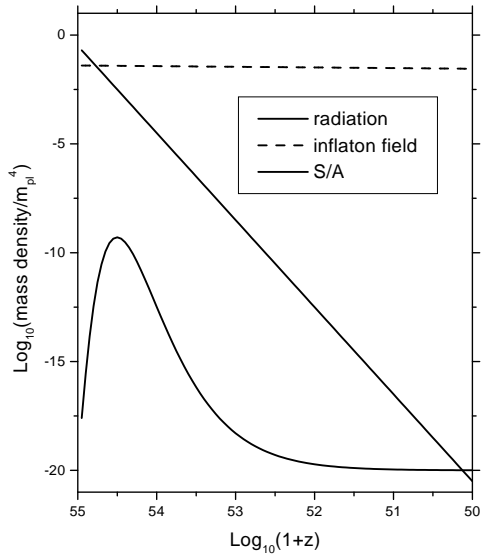}\\
\caption[graph1]{\label{graph1} This plot shows a detailed
evolution of the energy densities and the holographic bound at
early times. This shows how inflation appears to protect the
violation of the HP.}
\end{figure}

To solve the cosmological problems, inflation must last around
$60$ e-folds. This means $a_{end}=e^{60}a_{init}$. This growth
makes the curvature term irrelevant for the subsequent analysis,
although it was fundamental at
the beginning. For example, if we assume that inflation begins when $%
H^{2}\sim a^{-2}$ in Eq.(\ref{hubble}), then after the $60$ $e$-folding we
have in Eq.(\ref{hubble}) $a^{-2}\simeq e^{-120}H^{2}\ll H^{2}$, making the
universe very flat.

After this 60 e-folding of inflation, particle creation starts.
During this phase the formulae derived in section IV are valid,
and the expression for the HP bound has to be replaced by
\[
\frac{S}{A}=\sigma \frac{\chi _{H}}{a^{2}}\propto \frac{H^{-1}}{a^{1/8}},
\]
which decreases more slowly than in the previous phase. Because
the curvature term has fallen more than $50$ orders of magnitude
from the beginning of inflation, it seems difficult to repeat a
similar argument to explain the current accelerated expansion.
This possibility is discussed in the next section.

\section{Possible extra inflationary stages}

After the 60 $e$-folding of inflation, bound \ref{s/a} has fallen
almost 100 orders of magnitude making impossible the existence of
HP saturation for another period . However, we have seen that
after reheating the universe becomes radiation-dominated, and we
can consider a small relic of the inflation field, which enables
us to have another accelerated period. This possibility is open if
the inflaton energy density does not fall more than 120 orders of
magnitude during reheating.

Based on observations, we have some good evidence of acceleration
for $z<0.5 $ and some preliminary evidence of deceleration for
$z>0.5$ \cite{turner01}. If we assume that $\rho _{\varphi }\simeq
\rho _{m}$ at $z=0.5$, we can explain the current acceleration of
the universe using the same field that drove inflation. Because at
small redshift the curvature could be relevant, we found for
$z<0.5$ the field $\varphi $ makes the universe look flat,
although closed. The evolution through the matter domination to
the present is very similar to what is shown in Figure 1.

\section{Discussion}

I have developed a consistent observational model based on the
Holographic Principle which explains the role of inflation. It
uses the HP as a relevant principle for inflation and leads to a
scenario where it is possible to explain the current observation
of accelerated expansion. In this way it reduces the amount of
scalar fields needed to explain inflation and the dark energy. If
reheating is efficient enough to change the relative weight of
energy densities, it can also explain the coincidence problem. I
have not used a particular potential form, using instead the
inflaton equation of state as the relevant object of study. After
reheating, the universe becomes radiation-dominated at $z\sim 20
$, and the inflaton energy density remains constant during all
this period, making the model consistent with nucleosynthesis and
making it behave similarly to the $\Lambda$CDM model. Although our
model has to be tuned to get the inflation domination after $z
\sim 0.5 $, we think that its main quality -- that of connect
inflation and the dark energy with holographic properties of the
universe -- are of sufficient interest to deserve further study.

\section*{Acknowledgments}

The author wishes to thank S. del Campo and J. Saavedra for
valuable discussions. This work has been supported by FONDECYT
grant {\bf 3010017}.


\begin{references}
\bibitem{inf}  E.W.Kolb and M.S. Turner, {\it The Early Universe},
Addison-Wesley, Redwood City (1990); A. Linde, {\it Particle Physics and
Inflationary Cosmology}, Harwood, Chur (1990); J.A. Peacock, {\it %
Cosmological Physics}, Cambridge University Press, Cambridge (1999).

\bibitem{thooft}  G. 'tHooft, gr-qc/9310026, published in {\it %
Salam-festschrift: a collection of talks}, eds. A. Ali, J. Ellis and S.
Randjbar-Daemi (Worl Scientific).

\bibitem{string}  L. Susskind, J. Math. Phys. {\bf 36}, 6377
(1995); J.M. Maldacena, Adv. Theor. Math. Phys. {\bf 2}, 231
(1998); E. Witten, Adv. Theor. Math. Phys. {\bf 2}, 253 (1998).

\bibitem{bek} J.D. Bekenstein, Phys. Rev. D {\bf 23}, 287 (1981);
 Int. J. Theor. Phys. {\bf 28} 967 (1989).

\bibitem{FS}  W. Fischler and L. Susskind, hep-th/9806039.

\bibitem{holo}  D. Bak and S.-J. Rey, Class. Quant. Grav. {\bf 17,} L83
(2000) ; R. Bousso, JHEP {\bf 9907, }004 (1999) ; JHEP {\bf 9906, }028
(1999); Class.Quant.Grav. {\bf 17,} 997 (2000) ; R. Tavakol and G. Ellis,
Phys. Lett. {\bf B469}, 37 (1999).

\bibitem{EL}  R. Easther and D. Lowe, Phys. Rev. Lett. {\bf 82}, 4967
(1999); G. Veneziano, Phys. Lett. {\bf B454}, 22 (1999).

\bibitem{KL}  N. Kaloper and A. Linde, Phys. Rev. D {\bf \ 60}, 103509 (1999).

\bibitem{rama}  S. K. Rama, Phys. Lett. {\bf B457}, 268 (1999).

\bibitem{QTS}  Pierre Binetruy, Int. J. Theor.Phys. {\bf 39,} 1859 (2000).

\bibitem{closed}  P. de Bernadis et al., astro-ph/0011469; J.R. Bond, et
al., astro-ph/0011378; R. Stompor, et al., Astrophys. J. {\bf
561}, L7 (2001), astro-ph/0105062; X. Wang, et al., Phys. Rev. D
{\bf 65}, 123001 (2002), astro-ph/0105091; M. Douspis, et al.,
Astron. Astrophys. {\bf 379}, 1 (2001), astro-ph/0105129; P. de
Bernadis, et al., astro-ph/0105269.

\bibitem{PV}  P.J.E. Peebles and A. Vilenkin, Phys. Rev. D {\bf 59}, 06505
(1999).

\bibitem{KT}  M. Kamionkowski and N. Toumbas, Phys. Rev. Lett. {\bf 77}, 587
(1996); N. Cruz, S. del Campo, and R. Herrera, Phys. Rev. D {\bf 58}, 123504
(1998).

\bibitem{REHEAT}  A.D. Dolgov and A. Linde, Phys. Lett. {\bf B116}, 329
(1982); L.F. Abbott, E. Farhi, and M.B. Wise, Phys. Lett. {\bf B117}, 29
(1982); A. Albrecht, P.J. Steinhardt, M.S. Turner, and F. Wilczek, Phys.
Rev. Lett. {\bf 48}, 1437 (1982); ; see also section 7.9 of A. Linde, {\it %
Particle Physics and Inflationary Cosmology}, Harwood, Chur (1990).

\bibitem{turner85}  R.J. Scherrer and M.S. Turner, Phys. Rev. D {\bf 31},
681 (1985).

\bibitem{MthR}  J.H. Traschen and R.H. Brandenberger, Phys. Rev. D {\bf 42},
2491 (1990); L. Kofman, A. Linde, and A.A. Starobinsky, Phys. Rev. Lett.
{\bf 73}, 3195 (1994); Y. Shtanov, J. Traschen, and R. Brandenberger, Phys.
Rev. D {\bf 51}, 5438 (1995); L.A. Kofman, A.D. Linde, and A.A. Starobinsky,
Phys. Rev. D {\bf 56}, 3258 (1997).

\bibitem{next}  V\'{\i}ctor H. C\'{a}rdenas, in preparation.

\bibitem{turner01}  M.S. Turner and A.G. Riess, arXiv preprint
astro-ph/0106051.
\end{references}
\end{document}